\documentclass[11pt]{article}
\usepackage{fullpage,latexsym,epsfig,amssymb,amsmath}

\newcommand{\adj}[1]{{#1}^{\dagger}}
\newcommand{\tuple}[1]{\mathord{\left\langle {#1} \right\rangle}}

\newcommand{\ket}[1]{\mathord{\left| {#1} \right\rangle}}
\newcommand{\braket}[2]{\tuple{{#1}|{#2}}}

\renewcommand{\phi}{\varphi}
\newcommand{\map}[3]{{#1} : {#2} \rightarrow {#3}}
\newcommand{\cH}{{\cal H}}

\newcommand{\wt}{{\rm wt}}

\newcommand{\eqdf}{\mathrel{:=}}

\newcommand{\absval}[1]{{\left|{#1}\right|}}

\newcommand{\setof}[1]{\left\{{#1}\right\}}
\newcommand{\two}{{\setof{0,1}}}

\newcommand{\floor}[1]{{\left\lfloor{#1}\right\rfloor}}

\newcommand{\Xtot}[1]{{\Sigma X}}
\newcommand{\Ytot}[1]{{\Sigma Y}}
\newcommand{\Ztot}[1]{{\Sigma Z}}
\newcommand{\Mod}{{\rm Mod}}

\newcommand{\raiseop}{{J_{\mathord +}}}
\newcommand{\lowerop}{{J_{\mathord -}}}

\title{Implementing fanout, parity, and Mod gates via spin exchange
interactions}

\author{Stephen A. Fenner\thanks{Computer Science and Engineering
Department, Columbia, SC 29208 USA.  Email {\tt
$\{$fenner$|$zhang29$\}$@cse.sc.edu}.
This work was supported in part by the National Security Agency
(NSA) and Advanced Research and Development Activity (ARDA) under Army
Research Office (ARO) contract number DAAD~190210048.} \\
University of South Carolina
\and
Yong Zhang\footnotemark[1]
\\
University of South Carolina}

\date{\today}

\begin{document}

\bibliographystyle{hplain}

\maketitle

\begin{abstract}
We show that, for any $n > 0$, the Heisenberg interaction among $2n$
qubits (as spin-1/2 particles) can be used to exactly implement an
$n$-qubit parity gate, which is equivalent in constant depth to an
$n$-qubit fanout gate.  Either isotropic or nonisotropic versions of
the interaction can be used.  We generalize our basic results by
showing that any Hamiltonian (acting on suitably encoded logical
qubits), whose eigenvalues depend quadratically on the Hamming weight
of the logical qubit values, can be used to implement generalized
$\Mod_q$ gates for any $q\geq 2$.

This paper is a sequel to quant-ph/0309163, and resolves a question
left open in that paper.
\end{abstract}

\section{Introduction}

\begin{figure}
\begin{center}
\begin{picture}(0,0)%
\includegraphics{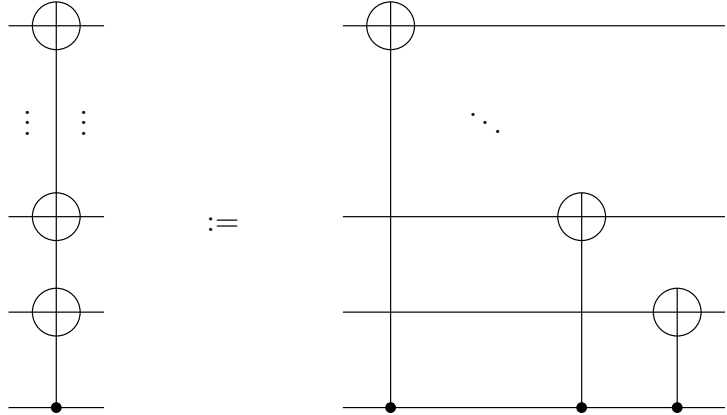}%
\end{picture}%
\setlength{\unitlength}{3947sp}%
\begingroup\makeatletter\ifx\SetFigFont\undefined%
\gdef\SetFigFont#1#2#3#4#5{%
  \reset@font\fontsize{#1}{#2pt}%
  \fontfamily{#3}\fontseries{#4}\fontshape{#5}%
  \selectfont}%
\fi\endgroup%
\begin{picture}(4524,2599)(2989,-3998)
\put(4351,-2836){\makebox(0,0)[b]{\smash{\SetFigFont{12}{14.4}{\rmdefault}{\mddefault}{\updefault}$:=$}}}
\put(3151,-2236){\makebox(0,0)[rb]{\smash{\SetFigFont{12}{14.4}{\rmdefault}{\mddefault}{\updefault}$\vdots$}}}
\put(3451,-2236){\makebox(0,0)[lb]{\smash{\SetFigFont{12}{14.4}{\rmdefault}{\mddefault}{\updefault}$\vdots$}}}
\put(6001,-2236){\makebox(0,0)[b]{\smash{\SetFigFont{12}{14.4}{\rmdefault}{\mddefault}{\updefault}$\ddots$}}}
\end{picture}
\caption{Definition of the fanout gate.}\label{fig:fanout-def}
\end{center}
\end{figure}

Let $\cH$ be the Hilbert space of $n$ qubits, where $n\geq 1$.  The
\emph{fanout operator} $\map{F_n}{\cH}{\cH}$, depicted in
Figure~\ref{fig:fanout-def},
copies the (classical) value of a single qubit to $n$ other qubits.
Unbounded fanout is usually taken for granted in models of classical
Boolean circuits, even when the fanin of gates is bounded.  They
cannot be taken for granted in quantum circuits, however, since
copying the value of a quantum bit to $n-1$ other bits requires
significant nonlocal interactions.  Unbounded fanout gates have been
shown to be a surprisingly powerful primitive for quantum computation,
allowing one to reduce the depth of a circuit computing, say, the
Quantum Fourier Transform (QFT) to essentially constant depth
\cite{HS:fanout}.  The quantum part of Shor's factoring algorithm can
thus be implemented in constant depth if unbounded fanout gates are
available.  This result about the power of fanout is especially
important because in most of the significant proposals for
implementing quantum circuits to date, long computations will surely
be difficult to maintain due to decoherence, current quantum error
correction techniques notwithstanding.  Shallow quantum circuits may
prove to be, at least in the short term, the only realistic model of
feasible quantum computation, and fanout gates would increase their
power significantly.

Without some quantum gate with unbounded width (arity), it is not
clear that any nontrivial decision problem can be computed by $o(\log
n)$-depth quantum circuits with bounded error.  This is certainly true
if we only allow a one-qubit output measurement (then the output can
only depend on $2^{o(\log n)}$ input qubits---see \cite{FFGHZ:fanout}
for a discussion), but it also seems to be true even if we allow a
computational-basis measurement of an arbitrary number of qubits at
the end \cite{FGHZ:constant-depth}.  Even if we allow unbounded AND
gates (generalized Toffoli gates), it is not clear what we can do in
sublogarithmic depth.  We do know that we cannot approximate fanout
gates this way \cite{FFGHZ:fanout}.

To summarize, fanout gates are an extremely useful, and perhaps
necessary, primitive for allowing small-depth quantum circuits to
solve useful problems.  Furthermore, implementing fanout with a
conventional quantum circuit requires logarithmic depth, even if
unbounded AND gates are allowed.  Therefore, implementing a fast
fanout gate will require an unconventional approach.

We provide such an approach here by showing that the fanout operator
arises easily by evolving qubits via a simple and well-studied
Hamiltonian, the \emph{spin-exchange} or \emph{Heisenberg}
interaction, together with a modest amount of encoding and decoding of
qubits (which only requires constant-width gates and constant depth).
Our results answer positively a question by I. L. Chuang, who asked
how certain forms of the Heisenberg interaction, which are
implementable in the laboratory, may be useful for quantum computation
\cite{Chuang:hamiltonian1,Chuang:hamiltonian2}.  In particular, we
show that the fanout gate on $n$ logical qubits can be achieved
exactly by encoding them into $2n$ physical qubits (each a spin-$1/2$
particle), then applying the Heisenberg interaction to the encoded
qubits.  The interaction need not be isotropic; both isotropic and
nonisotropic versions of the interaction work equally well.

In \cite{Fenner:fanout}, we showed that a variant of the Heisenberg
interaction, where the Hamiltonian is proportional to the square of
the $z$-component of the total spin, can implement parity easily
(without encoding).  When applied to three qubits, this interaction
yields an ``inversion on equality gate'' $I_=$, defined by
\[ I_=\ket{xyz} = \left\{ \begin{array}{ll}
-\ket{xyz} & \mbox{if $x=y=z$,} \\
\ket{xyz}  & \mbox{otherwise.}
\end{array} \right. \]
$I_=$ and single-qubit gates together form a universal set of gates.
Recently, implementation of $I_=$ as well as the three-qubit parity
and fanout gates in NMR using the above Hamiltonian has been reported
\cite{GDK:fanoutNMR}.  Our current paper affirmatively answers a
question left open in \cite{Fenner:fanout} as to whether parity/fanout
can be implemented using more common forms of the Heisenberg
interaction, involving $x$-, $y$-, and $z$-components of the total
spin.

\begin{figure}
\begin{center}
\begin{picture}(0,0)%
\includegraphics{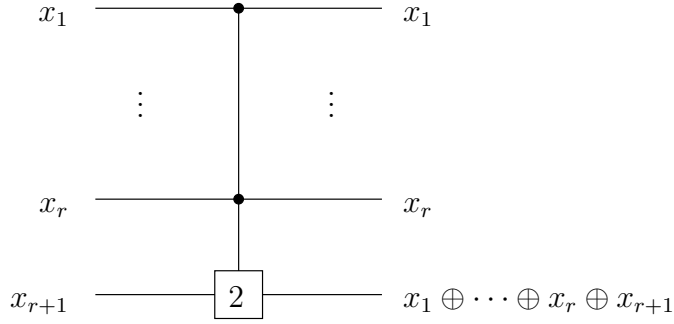}%
\end{picture}%
\setlength{\unitlength}{3947sp}%
\begingroup\makeatletter\ifx\SetFigFont\undefined%
\gdef\SetFigFont#1#2#3#4#5{%
  \reset@font\fontsize{#1}{#2pt}%
  \fontfamily{#3}\fontseries{#4}\fontshape{#5}%
  \selectfont}%
\fi\endgroup%
\begin{picture}(2100,2082)(2251,-2923)
\put(2701,-1636){\makebox(0,0)[b]{\smash{\SetFigFont{12}{14.4}{\rmdefault}{\mddefault}{\updefault}$\vdots$}}}
\put(3901,-1636){\makebox(0,0)[b]{\smash{\SetFigFont{12}{14.4}{\rmdefault}{\mddefault}{\updefault}$\vdots$}}}
\put(2251,-2236){\makebox(0,0)[rb]{\smash{\SetFigFont{12}{14.4}{\rmdefault}{\mddefault}{\updefault}$x_r$}}}
\put(4351,-2236){\makebox(0,0)[lb]{\smash{\SetFigFont{12}{14.4}{\rmdefault}{\mddefault}{\updefault}$x_r$}}}
\put(4351,-2836){\makebox(0,0)[lb]{\smash{\SetFigFont{12}{14.4}{\rmdefault}{\mddefault}{\updefault}$x_1\oplus \cdots \oplus x_r \oplus x_{r+1}$}}}
\put(4351,-1036){\makebox(0,0)[lb]{\smash{\SetFigFont{12}{14.4}{\rmdefault}{\mddefault}{\updefault}$x_1$}}}
\put(2251,-1036){\makebox(0,0)[rb]{\smash{\SetFigFont{12}{14.4}{\rmdefault}{\mddefault}{\updefault}$x_1$}}}
\put(2251,-2836){\makebox(0,0)[rb]{\smash{\SetFigFont{12}{14.4}{\rmdefault}{\mddefault}{\updefault}$x_{r+1}$}}}
\put(3301,-2836){\makebox(0,0)[b]{\smash{\SetFigFont{12}{14.4}{\rmdefault}{\mddefault}{\updefault}$2$}}}
\end{picture}
\caption{Definition of the parity gate.}\label{fig:parity-def}
\end{center}
\end{figure}

In Section~\ref{sec:prelims} we define the general Heisenberg
interaction between $n$ identical spins, as well as the special case
of interest to us.  In Section~\ref{sec:heisenberg} we give an
implementation of the $(r+1)$-bit parity gate, depicted in
Figure~\ref{fig:parity-def},
where $r = n/2$.  The fanout gate arises by placing Hadamard gates on
each qubit on both sides of the parity gate ($2(r+1)$ Hadamard gates
in all), and thus implementing parity is equivalent to implementing
fanout.  We also show in Section~\ref{sec:compressed} how different
qubit encoding schemes can reduce the ratio $n/r$ to be arbitrarily
close to one.  In Section~\ref{sec:generalize}, we generalize our
results in two ways: (1) any Hamiltonian whose eigenvalues depend
quadratically on the Hamming weight of the logical qubits can be used
to implement parity, and hence fanout, and (2) any such Hamiltonian
can implement generalized $\Mod_q$-gates directly for any $q\geq 2$.

\section{Preliminaries}
\label{sec:prelims}

The Heisenberg interaction describes the way particles in the same
general location affect each other by the magnetic moments arising
from their spin angular momenta.  Given a system of $m$ identical
labeled spins described by vector operators
$\vec{S}_1,\ldots,\vec{S}_m$, the Hamiltonian $E$ of the system is a
weighted sum of the energies of all the pairwise interactions, plus a
term for any external magnetic field (assumed to be in the
$z$-direction):
\begin{equation}\label{eqn:heisenberg}
E = - \sum_{i<j}J_{i,j} \vec{S}_i\cdot \vec{S}_j +
\alpha\sum_i (\vec{S}_i)_z,
\end{equation}
where the $J_{i,j}$ and $\alpha$ are constants.\footnote{The $J_{i,j}$
are usually assumed to be positive, appropriate for ferromagnetic
interactions which give the lowest energy when spins are aligned in
parallel.  The value of $\alpha$ is the product of the magnetic field
strength and the gyromagnetic ratio for the individual spins (see
\cite[\S 21.3]{Merzbacher:quantum} for example).}  In this paper, we
will show how this interaction can implement fanout in the special
case where all the $J_{i,j}$ are equal.  In this case, $E$ is related
to the squared magnitude of the total spin of the system.

We will assume here that physical qubits are implemented as spin-$1/2$
particles, with $\ket{0}$ being the spin-up state (in the positive
$z$-direction) and $\ket{1}$ being the spin-down state (in the
negative $z$-direction).  Given a system of $m$ qubits labeled
$1,\ldots,m$, we define
\begin{eqnarray*}
J_x & = & \frac{1}{2}\sum_{i=1}^m X_i \\
J_y & = & \frac{1}{2}\sum_{i=1}^m Y_i \\
J_z & = & \frac{1}{2}\sum_{i=1}^m Z_i
\end{eqnarray*}
where $X_i$, $Y_i$, and $Z_i$ are the three Pauli operators acting on
the $i$'th qubit.  $J_x$, $J_y$, and $J_z$ give the total spin in the
$x$-, $y$-, and $z$-directions, respectively.  The squared magnitude of
the total spin angular momentum of the system is given by the
observable
\[ J^2 = J_x^2 + J_y^2 + J_z^2. \]
Note that
\[ J^2 = \frac{3m}{4}I + \frac{1}{2}\sum_{1\leq i < j \leq m} (X_iX_j
+ Y_iY_j + Z_iZ_j) = \frac{3m}{4}I + \sum_{i<j} \vec{S}_i \cdot
\vec{S}_j, \] where $\vec{S}_i = \frac{1}{2}(X_i,Y_i,Z_i)$ is the
vector observable giving the spin of the $i$th qubit.  It is then
clear that, in the absence of an external magnetic field, $J^2$ is
linearly related to the energy $E$ above.  This is an isotropic
Heisenberg interaction.  To account for an external field in the
$z$-direction, we define, for any real $\alpha$,
\begin{equation}\label{eqn:H-def}
H_{\alpha} = - J^2 + \alpha J_z.
\end{equation}
This is the case of the Heisenberg interaction where all the $J_{i,j}$
are unity.\footnote{This does not lose generality, since constant
factors in the energy can be absorbed by adjusting the time of the
interaction.}  Our methods can also accommodate an extra term in the
Hamiltonian proportional to $J_z^2$ with no additional effort (see
Section~\ref{sec:heisenberg}), so we consider the more general
Hamiltonian
\begin{equation}\label{eqn:general-hamiltonian}
H_{\alpha,\beta} = - J^2 + \alpha J_z + \beta J_z^2
\end{equation}
for any real $\alpha$ and $\beta$ such that $\beta \neq 1$.  We will evolve
the system of $m$ qubits using $H_{\alpha,\beta}$ as the Hamiltonian.
A related Hamiltonian
$J_z^2$ is used in \cite{Fenner:fanout} to implement fanout; this is
an easy case, since each computational basis state is already an
eigenstate of $J_z^2$ and thus the implementation requires no encoding
of qubits.  Using the current Hamiltonian $H_{\alpha,\beta}$ is more
complicated and requires encoding logical qubits into groups of
physical qubits, so that the tensor product of all the physical qubits
encoding a logical basis state will be an eigenstate of
$H_{\alpha,\beta}$.

In the sequel, we choose units so that $\hbar = 1$.  If $A$ and $B$
are both vectors or both operators, we say, ``$A\propto B$'' to mean
that $A = e^{i\theta}B$ for some real $\theta$, that is, $A = B$ up to
an overall phase factor.  We use the same notation with individual
components of $A$ and $B$, meaning that the phase factor is
independent of which component we choose.
If $x\in\two^n$ is a bit
vector, we let $\wt(x)$ denote the Hamming weight of $x$, that is, the
number of $1$s in $x$.

\subsection{Spin States}

The properties of the $J$-operators are well-known.  See, for example,
B\"{o}hm \cite{Boehm:quantum}.  We will review the essential ones
here.  The commutation relations are $[J_x,J_y] = iJ_z$, and likewise
for the two other cyclic shifts of the indices.  $J^2$ commutes with
$J_z$, so one may choose an orthonormal basis of the $n$-qubit Hilbert
space $\cH$ that diagonalizes both simultaneously.  Eigenstates of
$J^2$ and $J_z$ are traditionally labeled as $\ket{j,m,\ell}$, where
$0\leq j \leq n/2$ and $-j \leq m \leq j$, with $n/2 - j$ and $j-m$
both integers.  We have $J_z\ket{j,m,\ell} = m\ket{j,m,\ell}$, and
$J^2\ket{j,m,\ell} = j(j+1)\ket{j,m,\ell}$.  The extra parameter $\ell$ is
used to give distinct labels to different basis vectors in degenerate
eigenspaces of $J^2$ and $J_z$.  These basis vectors can be chosen so that,
for any value $\ell$ that appears as the third label of some basis
vector, the basis vectors labeled by $\ell$ span an irreducible spin
representation, that is, a minimal subspace of $\cH$ invariant under
the action of $J_x$, $J_y$, and $J_z$.  This space will be spanned by
the basis vectors $\ket{j,-j,\ell},\ket{j,-j+1,\ell}, \ldots,
\ket{j,j-1,\ell},\ket{j,j,\ell}$, for some $j = j(\ell)$ depending
only on the label $\ell$, and is called a spin-$j$ representation.
Letting $\raiseop = J_x + iJ_y$ and $\lowerop = \adj{\raiseop} = J_x -
iJ_y$ be the usual raising and lowering operators, respectively, we
may adjust the phases of the basis vectors so that
\begin{eqnarray*}
\raiseop\ket{j,m,\ell} & = & \sqrt{j(j+1) - m(m+1)}\; \ket{j,m+1,\ell}, \\
\lowerop\ket{j,m,\ell} & = & \sqrt{j(j+1) - m(m-1)}\; \ket{j,m-1,\ell}.
\end{eqnarray*}
(This sets the relative phases of states within the representation,
but still allows the overall phase of the representation to be
adjusted relative to other representations.)

\subsection{Number of Spin Representations}
\label{sec:num-spin-reps}

An important fact that we will use later is that for each $j$, there
are exactly $k_{n,j} \eqdf {n \choose n/2 - j} - {n \choose n/2 - j -
1}$ many spin-$j$ representations in the decomposition of
$\cH$.\footnote{By convention, if $k<0$ then ${n \choose k} = 0$.}
One way to see this is as follows.  For any $j \geq 0$ such that $n/2
- j$ is an integer, let $\cH_j$ be the eigenspace of $J_z$ with
eigenvalue $j$ (if $j > n/2$, then $\cH_j$ has dimension zero).
Clearly, $\dim(\cH_j) = {n \choose n/2 - j}$, since $\cH_j$ is spanned
by all the computational basis vectors with Hamming weight $n/2 - j$.
The $\raiseop$ operator maps $\cH_j$ into $\cH_{j+1}$, and so its
kernel on $\cH_j$ has dimension at least $k_{n,j} > 0$, given above.
Now the space
$\ker(\raiseop)\cap\cH_j$ is spanned by the set of all states of the
form $\ket{j,j,\ell}$, so there are no less than $k_{n,j}$ distinct
values for $\ell$ occuring in the set, and each one labels a distinct
spin-$j$ representation.  Finally, since each spin-$j$ representation
has dimension $2j+1$ for all $j$, and since $\cH$ has $2^n$
dimensions, a simple counting argument shows that there can be no more
than $k_{n,j}$ many spin-$j$ representations, either.

It follows that there are a total of ${n \choose \floor{n/2}}$ many
spin representations in the decomposition of $\cH$.  Moreover, if $n$
is even, then it is easy to show that the representations are evenly
split between those where $n/2-j$ is even and those where $n/2-j$ is
odd: $\frac{1}{2} {n \choose n/2}$ representations for each.  This
fact will be used in Section~\ref{sec:compressed}.

\subsection{Spin States versus Computational Basis States}
\label{sec:spin-vs-comp}

Finally, we mention how some of the spin states relate to
computational basis states.  There is one spin-$n/2$ representation in
the decomposition, namely, the completely symmetric representation,
which is spanned by the states
\[ \ket{n/2,n/2 - k} = {n \choose k}^{-1/2} \sum_{\wt(x)=k} \ket{x} \]
for integer $k$ with $0\leq k \leq n$.  (This equation sets the
overall phase of the spin-$n/2$ representation.)  A key point in this
paper is to note that $\ket{\frac{n}{2},\frac{n}{2}} = \ket{0^n}$ is a
tensor product of single qubits.  This means that some spin states
involve little or no entanglement among qubits and thus can be
prepared using only reasonably local interactions.  More generally,
suppose we group some of the $n$ qubits into disjoint pairs
$(i_1,j_1),(i_2,j_2),\ldots,(i_p,j_p)$ for some $0\leq p \leq n/2$
(that is, $i_1,j_1,\ldots,i_p,j_p\in\setof{1,\ldots,n}$ and are all
pairwise distinct), then we form a state $\ket{\psi}$ by putting each
pair $(i_k,j_k)$ of qubits into the singlet state $\ket{0,0} =
(\ket{10} - \ket{01})/\sqrt{2}$ and each of the rest of the (unpaired)
qubits into the $\ket{0}$ state, i.e.,
\begin{equation}\label{eqn:pairing}
\ket{\psi} = \ket{0,0}_{i_1,j_1} \cdots \ket{0,0}_{i_p,j_p}
\ket{0^{n-2p}}_S = \ket{0,0}_{i_1,j_1} \cdots \ket{0,0}_{i_p,j_p}
\ket{j,j}_S,
\end{equation}
where $j = n/2 - p$, and $S$ is the set of all unpaired qubits.  Then
it is easy to check that $\ket{\psi}$ is an eigenstate of $J_z$ (with
eigenvalue $n/2 - p$) and is in $\ker(\raiseop)$.  It follows that
$\ket{\psi}$ is also an eigenstate of $J^2$, because $0 =
\lowerop\raiseop\ket{\psi} = (J^2 - J_z^2 - J_z)\ket{\psi}$.  Many of
the states $\ket{j,j,\ell}$ where $j = n/2 - p$ can be defined this
way, but not all---different choices of the $p$ pairs do not always
produce states that are orthogonal to each other, or even linearly
independent.

\section{Main Results}

\subsection{Parity Gate by Heisenberg Interactions}
\label{sec:heisenberg}

Recall the Hamiltonian $H_{\alpha,\beta}$ of
(\ref{eqn:general-hamiltonian}) on the space $\cH$ of $n$ qubits
labeled $1,\ldots,n$.  $H_{\alpha,\beta}$ commutes with $J^2$, so it
has eigenvectors $\ket{j,m,\ell}$ with respective eigenvalues $-j(j+1)
+ \alpha m + \beta m^2$.

\begin{figure}
\begin{center}
\begin{picture}(0,0)%
\includegraphics{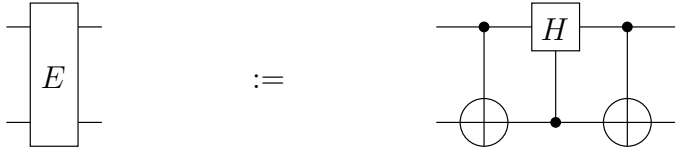}%
\end{picture}%
\setlength{\unitlength}{3947sp}%
\begingroup\makeatletter\ifx\SetFigFont\undefined%
\gdef\SetFigFont#1#2#3#4#5{%
  \reset@font\fontsize{#1}{#2pt}%
  \fontfamily{#3}\fontseries{#4}\fontshape{#5}%
  \selectfont}%
\fi\endgroup%
\begin{picture}(4224,924)(3889,-2623)
\put(5551,-2236){\makebox(0,0)[b]{\smash{\SetFigFont{12}{14.4}{\rmdefault}{\mddefault}{\updefault}$:=$}}}
\put(4201,-2236){\makebox(0,0)[b]{\smash{\SetFigFont{12}{14.4}{\rmdefault}{\mddefault}{\updefault}$E$}}}
\put(7351,-1936){\makebox(0,0)[b]{\smash{\SetFigFont{12}{14.4}{\rmdefault}{\mddefault}{\updefault}$H$}}}
\end{picture}
\caption{A two-qubit encoder.}\label{fig:encode}
\end{center}
\end{figure}

Let $x = x_1\cdots x_r$ be a vector of $r$ bits, where $r$ is some
number no greater than $n$.  We wish to encode the $r$-qubit
computational basis state $\ket{x}$ into an $n$-qubit eigenstate
$\ket{j,m,\ell}$ of $H_{\alpha,\beta}$ so that $j$ depends linearly on
$\wt(x)$, and we want to do this by using gates that act on as few
qubits as possible.
The easiest way to accomplish this is to have $n = 2r$ and create
encoded states of the form given by (\ref{eqn:pairing}).  One may
encode each input qubit (with an ancilla) into two qubits, sending
$\ket{00}$ to $\ket{0_L} \eqdf \ket{00}$ and sending $\ket{10}$ to
$\ket{1_L} \eqdf (\ket{10} - \ket{01})/\sqrt{2}$, the singlet state.
A simple circuit---one of many---for this is shown in
Figure~\ref{fig:encode},
which defines the encoding operator $E$\@.  (Since $\adj{E} = E$, we
will also use $E$ to decode.)  Clearly, there are other operators that
will work just as well, since $E$ is underdetermined.  We encode the
$i$th input with ancilla into the qubits $2i-1$ and $2i$ of $\cH$.
Thus $\ket{x_1\cdots x_r}$ maps to $\ket{x_L} \eqdf
\ket{x_{1,L}}\otimes \cdots \otimes \ket{x_{r,L}}$, and this state is
in the form of (\ref{eqn:pairing}), where the set $S$ consists of all
physical qubits encoding the $\ket{0_L}$ states, i.e., $S =
\bigcup_{x_i=0}\setof{2i-1,2i}$.  If $x \neq y$, then clearly
$\braket{x_L}{y_L} = 0$, so we can assume without loss of generality
that $\ket{x_L} = \ket{j(x),j(x),x}$, where $j(x) \eqdf n/2 - \wt(x)$,
and $x$ itself is used for the label.

Suppose that $\beta \neq 1$, and let
\begin{equation}\label{eqn:t-def}
t \eqdf \frac{\pi}{2\absval{\beta - 1}}.
\end{equation}
We let
\begin{equation}\label{eqn:U-def}
U \eqdf e^{-itH_{\alpha,\beta}},
\end{equation}
the unitary operator resulting from evolving the qubits with
$H_{\alpha,\beta}$ for time $t$.  For fixed input vector $x =
x_1\cdots x_r \in\two^r$, let $k = \wt(x)$, and for $b\in\two$ let
$x^b \eqdf x_1\cdots x_{r-1}b$.  To compute the parity of $k$ with a
quantum circuit on input $\ket{x}$, we first run qubit $r$ through a
Hadamard gate to produce the state
\[ \ket{\varphi_x} \eqdf \frac{\ket{x^0} + (-1)^{x_r}\ket{x^1}}{\sqrt{2}}. \]
Next, we encode each qubit as described above to obtain
\[ \ket{\varphi_{x,L}} \eqdf \frac{\ket{x^0_L} +
(-1)^{x_r}\ket{x^1_L}}{\sqrt{2}} = \frac{\ket{j_0,j_0,x^0} +
(-1)^{x_r}\ket{j_1,j_1,x^1}}{\sqrt{2}}, \]
where we have set $j_b \eqdf j(x^b) = n/2 - \wt(x^b)$.  Observing that
$j_1 = j_0 - 1$, we see that $\ket{\varphi_{x,L}}$ is a balanced
superposition of two spin states, one with even $j$ and the other with
odd $j$.  We now apply $U$ to $\ket{\varphi_{x,L}}$.  Noting that
$H_{\alpha,\beta}\ket{j,j,\ell} = ((\beta - 1)j^2 + (\alpha -
1)j)\ket{j,j,\ell}$ for any $j$, we set
\begin{equation}\label{eqn:gamma-def}
\gamma \eqdf \frac{\alpha - 1}{\beta - 1},
\end{equation}
and define
\begin{equation}\label{eqn:U-action}
\ket{\eta_{j,\ell}} \eqdf U\ket{j,j,\ell} = 
e^{-is\pi(j^2 + \gamma j)/2}\ket{j,j,\ell}
\end{equation}
for any $j$ and $\ell$, where for convenience, we are letting
\[ s \eqdf \frac{\beta-1}{\absval{\beta-1}} = \left\{ \begin{array}{ll}
1 & \mbox{if $\beta > 1$,} \\
-1 & \mbox{if $\beta < 1$.}
\end{array} \right. \]
We have
\begin{eqnarray*}
U\ket{\varphi_{x,L}}
& = & (\ket{\eta_{j_0,x^0}} +
(-1)^{x_r}\ket{\eta_{j_1,x^1}})/\sqrt{2} \\
& = & \left[\exp(-is\pi(j_0^2+\gamma j_0)/2)\ket{j_0,j_0,x^0}\right. \\
& & \mbox{ } + \left.(-1)^{x_r}\exp(-is\pi(j_1^2+\gamma
j_1)/2)\ket{j_1,j_1,x^1}\right]/\sqrt{2} \\
& = & \ket{x_{1,L}} \otimes \cdots \otimes \ket{x_{r-1,L}} \otimes
\ket{\Psi_x},
\end{eqnarray*}
where
\begin{eqnarray}
\ket{\Psi_x} & = & \frac{e^{-is\pi(j_0^2+\gamma
j_0)/2}\ket{0_L} + (-1)^{x_r}e^{-is\pi(j_1^2+\gamma
j_1)/2}\ket{1_L}}{\sqrt{2}} \label{eqn:Psi-first} \\
& = & \frac{e^{-is\pi(j_0^2 + \gamma j_0)/2}}{\sqrt{2}} \left[
\ket{0_L} + (-1)^{x_r} e^{-is\pi(-2j_0 + 1 - \gamma)/2} \ket{1_L}
\right] \\
& \propto & \frac{1}{\sqrt{2}}\left[ \ket{0_L} + (-1)^{j_0 + x_r}
e^{is\pi(\gamma - 1)/2} \ket{1_L} \right] \\
& = & \frac{1}{\sqrt{2}}\left[ \ket{0_L} + (-1)^{r + \wt(x)}
e^{is\pi(\gamma - 1)/2} \ket{1_L} \right] \label{eqn:Psi-last}
\end{eqnarray}
is the state of the $(2r-1)$st and $(2r)$th qubits.  (Recall that $j_1
= j_0 - 1$, and that $j_0 = \frac{n}{2} - \wt(x^0) = r - \wt(x) +
x_r$.)

For any $y = y_1\cdots y_r \in\two^r$, it is easy to see that if
$\wt(x)$ and $\wt(y)$ have opposite parity, then $\ket{\Psi_y}$ and
$\ket{\Psi_x}$ are orthogonal: it follows immediately from
(\ref{eqn:Psi-first}--\ref{eqn:Psi-last}) that
\[ 2\braket{\Psi_y}{\Psi_x} \propto 1 + (-1)^{\wt(x)+\wt(y)}, \]
which is zero if $\wt(x)$ and $\wt(y)$ have opposite parity.  This
analysis shows that we have isolated the parity information in the
$r$th logical qubit.  We decode \emph{just} the two physical qubits
corresponding to this qubit to obtain the state
\begin{eqnarray*}
E\ket{\Psi_x} & = & \frac{e^{-is\pi(j_0^2 + \gamma j_0)/2}}{\sqrt{2}}
\left[ \ket{00} + (-1)^{r + \wt(x)} e^{is\pi(\gamma - 1)/2} \ket{10}
\right] \\
& \propto & \frac{1}{\sqrt{2}} \left[ \ket{0} + (-1)^{r + \wt(x)}
e^{is\pi(\gamma - 1)/2} \ket{1} \right] \otimes \ket{0}.
\end{eqnarray*}
The second of these qubits is the restored ancilla.  We then apply two
gates, $V$ followed by $H$, to the first qubit, where
\[ V \eqdf \left[ \begin{array}{cc} 1 & 0 \\ 0 &
e^{-is\pi(2r + \gamma - 1)/2} \end{array}\right], \]
and $H$ is the Hadamard transform.  This yields $\ket{\wt(x) \bmod 2}$
as the state of the first qubit, up to some unconditional phase
factor.

\begin{figure}
\begin{center}
\begin{picture}(0,0)%
\includegraphics{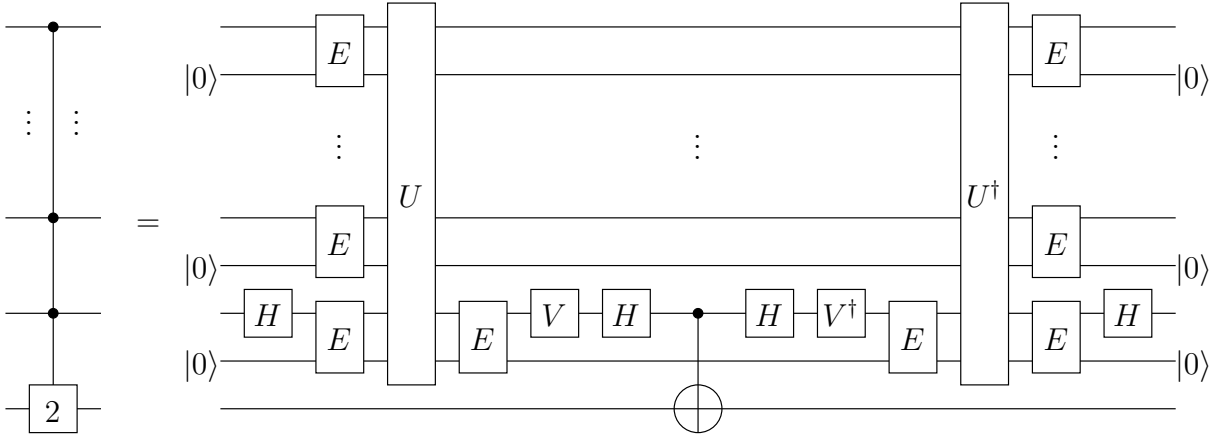}%
\end{picture}%
\setlength{\unitlength}{3947sp}%
\begingroup\makeatletter\ifx\SetFigFont\undefined%
\gdef\SetFigFont#1#2#3#4#5{%
  \reset@font\fontsize{#1}{#2pt}%
  \fontfamily{#3}\fontseries{#4}\fontshape{#5}%
  \selectfont}%
\fi\endgroup%
\begin{picture}(7374,2724)(2839,-4723)
\put(4201,-2536){\makebox(0,0)[rb]{\smash{\SetFigFont{12}{14.4}{\rmdefault}{\mddefault}{\updefault}$\ket{0}$}}}
\put(4201,-3736){\makebox(0,0)[rb]{\smash{\SetFigFont{12}{14.4}{\rmdefault}{\mddefault}{\updefault}$\ket{0}$}}}
\put(4201,-4336){\makebox(0,0)[rb]{\smash{\SetFigFont{12}{14.4}{\rmdefault}{\mddefault}{\updefault}$\ket{0}$}}}
\put(10201,-4336){\makebox(0,0)[lb]{\smash{\SetFigFont{12}{14.4}{\rmdefault}{\mddefault}{\updefault}$\ket{0}$}}}
\put(10201,-3736){\makebox(0,0)[lb]{\smash{\SetFigFont{12}{14.4}{\rmdefault}{\mddefault}{\updefault}$\ket{0}$}}}
\put(10201,-2536){\makebox(0,0)[lb]{\smash{\SetFigFont{12}{14.4}{\rmdefault}{\mddefault}{\updefault}$\ket{0}$}}}
\put(7201,-2986){\makebox(0,0)[b]{\smash{\SetFigFont{12}{14.4}{\rmdefault}{\mddefault}{\updefault}$\vdots$}}}
\put(3751,-3436){\makebox(0,0)[b]{\smash{\SetFigFont{12}{14.4}{\rmdefault}{\mddefault}{\updefault}$=$}}}
\put(3001,-2836){\makebox(0,0)[b]{\smash{\SetFigFont{12}{14.4}{\rmdefault}{\mddefault}{\updefault}$\vdots$}}}
\put(3301,-2836){\makebox(0,0)[b]{\smash{\SetFigFont{12}{14.4}{\rmdefault}{\mddefault}{\updefault}$\vdots$}}}
\put(6751,-4036){\makebox(0,0)[b]{\smash{\SetFigFont{12}{14.4}{\rmdefault}{\mddefault}{\updefault}$H$}}}
\put(7651,-4036){\makebox(0,0)[b]{\smash{\SetFigFont{12}{14.4}{\rmdefault}{\mddefault}{\updefault}$H$}}}
\put(5851,-4186){\makebox(0,0)[b]{\smash{\SetFigFont{12}{14.4}{\rmdefault}{\mddefault}{\updefault}$E$}}}
\put(8551,-4186){\makebox(0,0)[b]{\smash{\SetFigFont{12}{14.4}{\rmdefault}{\mddefault}{\updefault}$E$}}}
\put(5401,-3286){\makebox(0,0)[b]{\smash{\SetFigFont{12}{14.4}{\rmdefault}{\mddefault}{\updefault}$U$}}}
\put(4951,-2386){\makebox(0,0)[b]{\smash{\SetFigFont{12}{14.4}{\rmdefault}{\mddefault}{\updefault}$E$}}}
\put(4951,-3586){\makebox(0,0)[b]{\smash{\SetFigFont{12}{14.4}{\rmdefault}{\mddefault}{\updefault}$E$}}}
\put(4951,-4186){\makebox(0,0)[b]{\smash{\SetFigFont{12}{14.4}{\rmdefault}{\mddefault}{\updefault}$E$}}}
\put(4951,-2986){\makebox(0,0)[b]{\smash{\SetFigFont{12}{14.4}{\rmdefault}{\mddefault}{\updefault}$\vdots$}}}
\put(9001,-3286){\makebox(0,0)[b]{\smash{\SetFigFont{12}{14.4}{\rmdefault}{\mddefault}{\updefault}$\adj{U}$}}}
\put(9451,-2386){\makebox(0,0)[b]{\smash{\SetFigFont{12}{14.4}{\rmdefault}{\mddefault}{\updefault}$E$}}}
\put(9451,-3586){\makebox(0,0)[b]{\smash{\SetFigFont{12}{14.4}{\rmdefault}{\mddefault}{\updefault}$E$}}}
\put(9451,-4186){\makebox(0,0)[b]{\smash{\SetFigFont{12}{14.4}{\rmdefault}{\mddefault}{\updefault}$E$}}}
\put(9451,-2986){\makebox(0,0)[b]{\smash{\SetFigFont{12}{14.4}{\rmdefault}{\mddefault}{\updefault}$\vdots$}}}
\put(6301,-4036){\makebox(0,0)[b]{\smash{\SetFigFont{12}{14.4}{\rmdefault}{\mddefault}{\updefault}$V$}}}
\put(8101,-4036){\makebox(0,0)[b]{\smash{\SetFigFont{12}{14.4}{\rmdefault}{\mddefault}{\updefault}$\adj{V}$}}}
\put(4501,-4036){\makebox(0,0)[b]{\smash{\SetFigFont{12}{14.4}{\rmdefault}{\mddefault}{\updefault}$H$}}}
\put(9901,-4036){\makebox(0,0)[b]{\smash{\SetFigFont{12}{14.4}{\rmdefault}{\mddefault}{\updefault}$H$}}}
\put(3151,-4636){\makebox(0,0)[b]{\smash{\SetFigFont{12}{14.4}{\rmdefault}{\mddefault}{\updefault}$2$}}}
\end{picture}
\caption{Circuit to implement parity with Heisenberg
interactions.}\label{fig:parity-circuit}
\end{center}
\end{figure}

If we do not mind the extra phase factor, which only depends on
$\gamma$, $r$, $s$, and $\wt(x_0)$, then we can simply use $E$ to decode
the other pairs of physical qubits, and we thus obtain a circuit that
computes parity.  To cleanly and exactly match the parity gate defined
in Figure~\ref{fig:parity-def}, however, we may first copy the parity
information onto a fresh qubit, then undo the previous computation.
The circuit for the latter operation is shown in
Figure~\ref{fig:parity-circuit}.
The gate on the left in Figure~\ref{fig:parity-circuit} is an
$(r+1)$-qubit parity gate.

\paragraph{Remark.}  The circuit of Figure~\ref{fig:parity-circuit}
uses the $\adj{U}$ gate.  Any unitary gate $U'$ that agrees with
$\adj{U}$ on the subspace $\cH'$ of $\cH$ spanned by vectors of the
form $\ket{j,j,\ell}$ can substitute for $\adj{U}$ to implement parity
exactly.  We would like to implement $U'$ by evolving the same
Hamiltonian $H_{\alpha,\beta}$ for some positive length of time.  We
can do this if there is a $u > 0$ such that $e^{-iuH_{\alpha,\beta}}$
fixes all vectors in $\cH'$.  Then we can implement $U'$ by evolving
$H_{\alpha,\beta}$ for time $ku-t$ where $t$ is given by
(\ref{eqn:t-def}) and $k$ is some integer such that $ku \geq t$, i.e.,
$U' \eqdf e^{-i(ku-t)H_{\alpha,\beta}}$.  By (\ref{eqn:U-action}), it
can be shown that such a $u$ exists if and only if the $\gamma$ of
(\ref{eqn:gamma-def}) is rational.  For arbitrary real $\gamma$, we
can still implement $U'$ by evolving the altered Hamiltonian
$H_{\alpha',\beta'}$ for some suitable $\alpha',\beta'$ and time $t'$.
Assuming $\beta' \neq 1$ and letting $\gamma' \eqdf
\frac{\alpha'-1}{\beta'-1}$, it follows from (\ref{eqn:U-action}) by a
straightforward argument that for $t' \geq 0$, the operator $U' \eqdf
e^{-it'H_{\alpha',\beta'}}$ is a suitable replacement for $\adj{U}$
(i.e., $U'U$ fixes all vectors in $\cH'$) if and only if there is an
integer $\ell$ such that (i) $(2\ell+1)(\gamma'+1)+s(\gamma+1)$ is an
integer multiple of four, (ii) $2\ell+1$ and $\beta' - 1$ have the
same sign, and (iii) $t' = \frac{\pi(2\ell+1)}{2(\beta'-1)}$.

\subsection{More Compressed Encodings}
\label{sec:compressed}

\begin{figure}
\begin{center}
\begin{picture}(0,0)%
\includegraphics{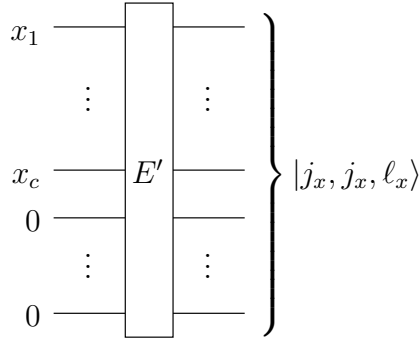}%
\end{picture}%
\setlength{\unitlength}{3947sp}%
\begingroup\makeatletter\ifx\SetFigFont\undefined%
\gdef\SetFigFont#1#2#3#4#5{%
  \reset@font\fontsize{#1}{#2pt}%
  \fontfamily{#3}\fontseries{#4}\fontshape{#5}%
  \selectfont}%
\fi\endgroup%
\begin{picture}(1350,2124)(3526,-3223)
\put(4201,-2236){\makebox(0,0)[b]{\smash{\SetFigFont{12}{14.4}{\rmdefault}{\mddefault}{\updefault}$E'$}}}
\put(3826,-2836){\makebox(0,0)[b]{\smash{\SetFigFont{12}{14.4}{\rmdefault}{\mddefault}{\updefault}$\vdots$}}}
\put(4576,-2836){\makebox(0,0)[b]{\smash{\SetFigFont{12}{14.4}{\rmdefault}{\mddefault}{\updefault}$\vdots$}}}
\put(3826,-1786){\makebox(0,0)[b]{\smash{\SetFigFont{12}{14.4}{\rmdefault}{\mddefault}{\updefault}$\vdots$}}}
\put(4576,-1786){\makebox(0,0)[b]{\smash{\SetFigFont{12}{14.4}{\rmdefault}{\mddefault}{\updefault}$\vdots$}}}
\put(3526,-1336){\makebox(0,0)[rb]{\smash{\SetFigFont{12}{14.4}{\rmdefault}{\mddefault}{\updefault}$x_1$}}}
\put(3526,-2236){\makebox(0,0)[rb]{\smash{\SetFigFont{12}{14.4}{\rmdefault}{\mddefault}{\updefault}$x_c$}}}
\put(3526,-2536){\makebox(0,0)[rb]{\smash{\SetFigFont{12}{14.4}{\rmdefault}{\mddefault}{\updefault}$0$}}}
\put(3526,-3136){\makebox(0,0)[rb]{\smash{\SetFigFont{12}{14.4}{\rmdefault}{\mddefault}{\updefault}$0$}}}
\put(4876,-2236){\makebox(0,0)[lb]{\smash{\SetFigFont{12}{14.4}{\rmdefault}{\mddefault}{\updefault}$\left.\phantom{\rule{0in}{0.875in}}\right\}\ket{j_x,j_x,\ell_x}$}}}
\end{picture}
\caption{A gate that encodes $c$ logical qubits $x = x_1\cdots x_c$
into $d$ physical qubits in state
$\ket{j_x,j_x,\ell_x}$.}\label{fig:group-encoder}
\end{center}
\end{figure}

In the previous section we encoded each logical qubit into two
physical qubits before applying $U$.  By encoding groups of logical
qubits, we can reduce the physical-to-logical qubit number ratio as
close as we want to unity.  Fix an integer $c>0$, and let $d$ be the
smallest even integer such that ${d \choose d/2} \geq 2^c$.  We can
compute parity as before by dividing the logical qubits into groups of
$c$ qubits each (assume for convenience that $r$ is a multiple of
$c$), and encoding each group into a group of $d$ physical qubits,
yielding a ratio of $d/c$.  Let $E'$ be such an encoder, depicted in
Figure~\ref{fig:group-encoder}.
Our only requirement for $E'$ is that it map each state
$\ket{x0^{d-c}}$, where $x\in\two^c$, into a state of the form
$\ket{j_x,j_x,\ell_x}$ where $j_x$ has the same parity as $d/2 -
\wt(x)$ and $\ell_x \neq \ell_y$ if $x\neq y$.  By our discussion in
Section~\ref{sec:num-spin-reps}, there are enough spin representations
on $d$ qubits to allow this, so such an $E'$ exists.

Now by the considerations of Section~\ref{sec:spin-vs-comp}, we see
that each input basis state $\ket{x}$ with $x\in\two^r$ is thus
encoded into a superposition of spin states $a_1\ket{j_1,j_1,\ell_1} +
a_2\ket{j_2,j_2,\ell_2} + \ldots\,$, where all the $j_i$ are integers
with parity equal to that of $n/2 - \wt(x)$.  We then see by linearity
that we can simulate the parity gate with a circuit identical to that
shown in Figure~\ref{fig:parity-circuit}, except that $E$ is replaced
with $E'$ or $\adj{E'}$ as appropriate, and each encoding group has
$d$ physical qubits.

Since ${d \choose d/2} \doteq 2^d/\sqrt{d}$, we see that
\[ \frac{d}{c} = \frac{d}{d - \frac{1}{2}\log_2 d} + O(1), \]
which shows the trade-off between the size of $E'$ and the ratio
$d/c$.

Encoding with an odd number of physical qubits per group is also
possible, and may sometimes lead to a slightly better trade-off.  For
example, there are enough spin representations on five physical qubits
to encode three logical qubits.

\section{Generalized Mod Gates from Any Quadratic Hamiltonian}
\label{sec:generalize}

In this section, we show how to implement a $\Mod_q$ gate directly,
for any $q\geq 2$, using any Hamiltonian whose eigenvalues depend
quadratically on the Hamming weights of the inputs.  More
specifically, we assume a Hamiltonian $G_n$ acting on $n$ qubits, real
constants $a_n,b_n,c_n$ with $a_n > 0$, and an encoding procedure
$E$ such that, for any computational basis state $\ket{x}$ over an
appropriate number of qubits, $E(\ket{x}\ket{00\cdots 0})$ is an
eigenstate of $G_n$ with eigenvalue $a_nw^2+b_nw+c_n$, where $w =
\wt(x)$ and $\ket{00\cdots 0}$ is some ancilla state.  Under these
assumptions, we construct circuits implementing $\Mod_q$ gates for any
constant $q\geq 2$, using evolution under $G_n$.  It is already known
that the $\Mod_q$ gates for all $q\geq 2$ are constant-depth
equivalent to each other \cite{GHMP:QAC}, so in effect, we already
have a Hamiltonian simulation of any $\Mod_q$ gate, via our
implementation of the parity ($\Mod_2$) gate and the simulation in
\cite{GHMP:QAC}.  Our approach here is much more direct, however.
Furthermore, it is only marginally more difficult conceptually to
generalize our simulation to all $q$, rather than just $q=2$.  Our
present development also subsumes the results in \cite{Fenner:fanout},
where we implemented parity ($q=2$) using the Hamiltonian $J_z^2$.

\begin{figure}
\begin{center}
\begin{picture}(0,0)%
\includegraphics{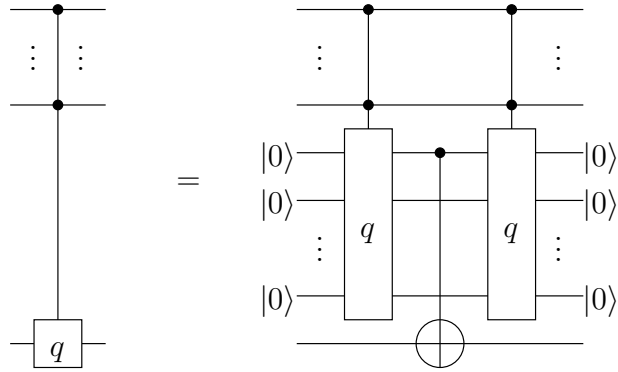}%
\end{picture}%
\setlength{\unitlength}{3947sp}%
\begingroup\makeatletter\ifx\SetFigFont\undefined%
\gdef\SetFigFont#1#2#3#4#5{%
  \reset@font\fontsize{#1}{#2pt}%
  \fontfamily{#3}\fontseries{#4}\fontshape{#5}%
  \selectfont}%
\fi\endgroup%
\begin{picture}(3624,2299)(2389,-5623)
\put(3526,-4486){\makebox(0,0)[b]{\smash{\SetFigFont{12}{14.4}{\rmdefault}{\mddefault}{\updefault}$=$}}}
\put(4351,-4936){\makebox(0,0)[b]{\smash{\SetFigFont{12}{14.4}{\rmdefault}{\mddefault}{\updefault}$\vdots$}}}
\put(4351,-3736){\makebox(0,0)[b]{\smash{\SetFigFont{12}{14.4}{\rmdefault}{\mddefault}{\updefault}$\vdots$}}}
\put(5851,-3736){\makebox(0,0)[b]{\smash{\SetFigFont{12}{14.4}{\rmdefault}{\mddefault}{\updefault}$\vdots$}}}
\put(5851,-4936){\makebox(0,0)[b]{\smash{\SetFigFont{12}{14.4}{\rmdefault}{\mddefault}{\updefault}$\vdots$}}}
\put(2851,-3736){\makebox(0,0)[b]{\smash{\SetFigFont{12}{14.4}{\rmdefault}{\mddefault}{\updefault}$\vdots$}}}
\put(2551,-3736){\makebox(0,0)[b]{\smash{\SetFigFont{12}{14.4}{\rmdefault}{\mddefault}{\updefault}$\vdots$}}}
\put(4201,-4636){\makebox(0,0)[rb]{\smash{\SetFigFont{12}{14.4}{\rmdefault}{\mddefault}{\updefault}$\ket{0}$}}}
\put(4201,-5236){\makebox(0,0)[rb]{\smash{\SetFigFont{12}{14.4}{\rmdefault}{\mddefault}{\updefault}$\ket{0}$}}}
\put(4201,-4336){\makebox(0,0)[rb]{\smash{\SetFigFont{12}{14.4}{\rmdefault}{\mddefault}{\updefault}$\ket{0}$}}}
\put(6001,-4336){\makebox(0,0)[lb]{\smash{\SetFigFont{12}{14.4}{\rmdefault}{\mddefault}{\updefault}$\ket{0}$}}}
\put(6001,-4636){\makebox(0,0)[lb]{\smash{\SetFigFont{12}{14.4}{\rmdefault}{\mddefault}{\updefault}$\ket{0}$}}}
\put(6001,-5236){\makebox(0,0)[lb]{\smash{\SetFigFont{12}{14.4}{\rmdefault}{\mddefault}{\updefault}$\ket{0}$}}}
\put(4651,-4786){\makebox(0,0)[b]{\smash{\SetFigFont{12}{14.4}{\rmdefault}{\mddefault}{\updefault}$q$}}}
\put(5551,-4786){\makebox(0,0)[b]{\smash{\SetFigFont{12}{14.4}{\rmdefault}{\mddefault}{\updefault}$q$}}}
\put(2701,-5536){\makebox(0,0)[b]{\smash{\SetFigFont{12}{14.4}{\rmdefault}{\mddefault}{\updefault}$q$}}}
\end{picture}
\caption{Simulating a standard $\Mod_q$ gate using generalized
$\Mod_q$ gates.  There are $r$ control qubits, and the ancill\ae\ on
the right are the qubits labeled
$t_1,t_2,\ldots,t_{q-1}$, i.e., the target qubits of the generalized
$\Mod_q$ gates.}\label{fig:mod-gates}
\end{center}
\end{figure}

Fix $q\geq 2$.  We consider $q$ to be constant.  The $\Mod_q$ gate is
a classical gate that acts on $r$ control bits and a target bit.  The
target bit is flipped iff the Hamming weight of the control bits is
not a multiple of $q$.  We will actually simulate a more powerful
version of this gate, the \emph{generalized $\Mod_q$ gate}, which has
$r$ control bits and $q-1$ target bits $t_1,\ldots,t_{q-1}$.  If $w$
is the Hamming weight of the control bits, then the target bits
$t_1,\ldots,t_i$ are all flipped, where $i = w \bmod q$, and the other
target bits are left alone.  Figure~\ref{fig:mod-gates}
shows how to simulate a (standard) $\Mod_q$
gate with a circuit using two generalized $\Mod_q$ gates and a CNOT
gate.

\begin{figure}
\begin{center}
\begin{picture}(0,0)%
\includegraphics{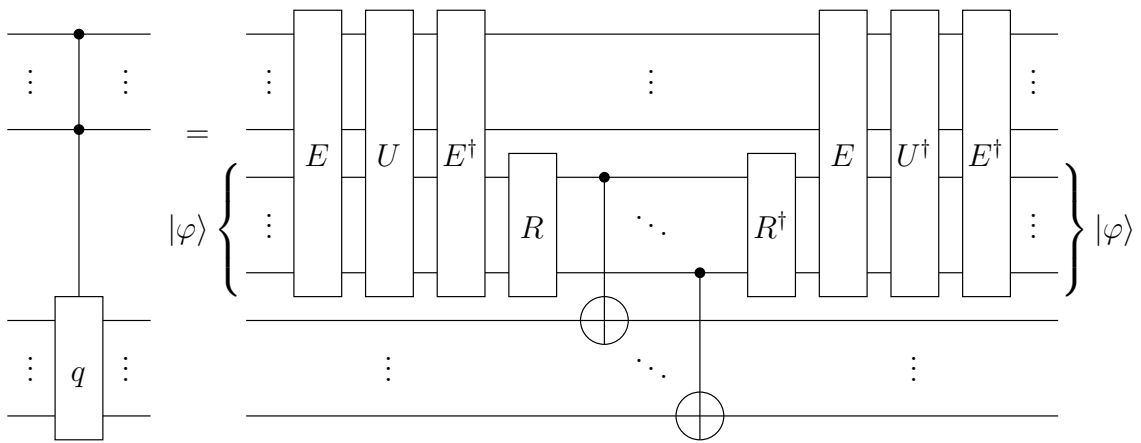}%
\end{picture}%
\setlength{\unitlength}{3947sp}%
\begingroup\makeatletter\ifx\SetFigFont\undefined%
\gdef\SetFigFont#1#2#3#4#5{%
  \reset@font\fontsize{#1}{#2pt}%
  \fontfamily{#3}\fontseries{#4}\fontshape{#5}%
  \selectfont}%
\fi\endgroup%
\begin{picture}(6624,2724)(3139,-4723)
\put(4351,-2836){\makebox(0,0)[b]{\smash{\SetFigFont{12}{14.4}{\rmdefault}{\mddefault}{\updefault}$=$}}}
\put(3301,-4336){\makebox(0,0)[b]{\smash{\SetFigFont{12}{14.4}{\rmdefault}{\mddefault}{\updefault}$\vdots$}}}
\put(3901,-4336){\makebox(0,0)[b]{\smash{\SetFigFont{12}{14.4}{\rmdefault}{\mddefault}{\updefault}$\vdots$}}}
\put(3901,-2536){\makebox(0,0)[b]{\smash{\SetFigFont{12}{14.4}{\rmdefault}{\mddefault}{\updefault}$\vdots$}}}
\put(3301,-2536){\makebox(0,0)[b]{\smash{\SetFigFont{12}{14.4}{\rmdefault}{\mddefault}{\updefault}$\vdots$}}}
\put(7201,-3436){\makebox(0,0)[b]{\smash{\SetFigFont{12}{14.4}{\rmdefault}{\mddefault}{\updefault}$\ddots$}}}
\put(7201,-4336){\makebox(0,0)[b]{\smash{\SetFigFont{12}{14.4}{\rmdefault}{\mddefault}{\updefault}$\ddots$}}}
\put(7201,-2536){\makebox(0,0)[b]{\smash{\SetFigFont{12}{14.4}{\rmdefault}{\mddefault}{\updefault}$\vdots$}}}
\put(5551,-4336){\makebox(0,0)[b]{\smash{\SetFigFont{12}{14.4}{\rmdefault}{\mddefault}{\updefault}$\vdots$}}}
\put(8851,-4336){\makebox(0,0)[b]{\smash{\SetFigFont{12}{14.4}{\rmdefault}{\mddefault}{\updefault}$\vdots$}}}
\put(4801,-2536){\makebox(0,0)[b]{\smash{\SetFigFont{12}{14.4}{\rmdefault}{\mddefault}{\updefault}$\vdots$}}}
\put(9601,-2536){\makebox(0,0)[b]{\smash{\SetFigFont{12}{14.4}{\rmdefault}{\mddefault}{\updefault}$\vdots$}}}
\put(4651,-3436){\makebox(0,0)[rb]{\smash{\SetFigFont{12}{14.4}{\rmdefault}{\mddefault}{\updefault}$\ket{\varphi}\left\{\rule{0in}{0.375in}\right.$}}}
\put(9751,-3436){\makebox(0,0)[lb]{\smash{\SetFigFont{12}{14.4}{\rmdefault}{\mddefault}{\updefault}$\left.\rule{0in}{0.375in}\right\}\ket{\varphi}$}}}
\put(4801,-3436){\makebox(0,0)[b]{\smash{\SetFigFont{12}{14.4}{\rmdefault}{\mddefault}{\updefault}$\vdots$}}}
\put(9601,-3436){\makebox(0,0)[b]{\smash{\SetFigFont{12}{14.4}{\rmdefault}{\mddefault}{\updefault}$\vdots$}}}
\put(6001,-2986){\makebox(0,0)[b]{\smash{\SetFigFont{12}{14.4}{\rmdefault}{\mddefault}{\updefault}$\adj{E}$}}}
\put(5551,-2986){\makebox(0,0)[b]{\smash{\SetFigFont{12}{14.4}{\rmdefault}{\mddefault}{\updefault}$U$}}}
\put(5101,-2986){\makebox(0,0)[b]{\smash{\SetFigFont{12}{14.4}{\rmdefault}{\mddefault}{\updefault}$E$}}}
\put(8401,-2986){\makebox(0,0)[b]{\smash{\SetFigFont{12}{14.4}{\rmdefault}{\mddefault}{\updefault}$E$}}}
\put(8851,-2986){\makebox(0,0)[b]{\smash{\SetFigFont{12}{14.4}{\rmdefault}{\mddefault}{\updefault}$\adj{U}$}}}
\put(9301,-2986){\makebox(0,0)[b]{\smash{\SetFigFont{12}{14.4}{\rmdefault}{\mddefault}{\updefault}$\adj{E}$}}}
\put(3601,-4336){\makebox(0,0)[b]{\smash{\SetFigFont{12}{14.4}{\rmdefault}{\mddefault}{\updefault}$q$}}}
\put(6451,-3436){\makebox(0,0)[b]{\smash{\SetFigFont{12}{14.4}{\rmdefault}{\mddefault}{\updefault}$R$}}}
\put(7951,-3436){\makebox(0,0)[b]{\smash{\SetFigFont{12}{14.4}{\rmdefault}{\mddefault}{\updefault}$\adj{R}$}}}
\end{picture}
\caption{Implementing a generalized $\Mod_q$ gate using the Hamiltonian $G_n$.  Here,
$U = e^{-itG_n}$ for an appropriate $t>0$, and $R$ is described below.  Any extra ancill\ae\ used
by the encoder $E$ are not shown.}\label{fig:mod-circuit}
\end{center}
\end{figure}

We use some $G_n$ to implement the generalized $\Mod_q$ gate via the
circuit shown in Figure~\ref{fig:mod-circuit}.
Given an initial basis state $\ket{x} = \ket{x_1\cdots x_r}$ of the
control qubits, we first prepare $q-1$ ancilla qubits into a state
\[ \ket{\varphi} \eqdf \sum_{j=0}^{q-1} c_j \ket{1^j0^{q-1-j}}, \]
where the $c_j$ are any fixed scalars such that $|c_j| = 1/\sqrt{q}$
(we may take $c_j = 1/\sqrt{q}$ for all $j$, for example).  Note that
$\ket{x}\ket{\varphi}$ is a superposition of basis states with
respective Hamming weights $\wt(x),\wt(x)+1,\ldots,\wt(x)+q-1$.  By
assumption, we have an encoder $E$ that maps each computational basis
state $\ket{y}$ with $y\in\two^{r+q-1}$ (possibly with additional
ancill\ae) to a state $\ket{y_L}$ over some number $n$ of qubits such
that
\[ G_n\ket{y_L} = (a_n\wt(y)^2 + b_n\wt(y) + c_n)\ket{y_L}. \]
Thus $E$ maps $\ket{x}\ket{\varphi}$ to the state
\[ \ket{\psi_{x,L}} \eqdf \sum_{j=0}^{q-1} c_j
\ket{(x1^j0^{q-1-j})_L}. \]
Next we apply $U \eqdf e^{-itG_n}$ to $\ket{\psi_{x,L}}$, where
$t = \frac{\pi k}{qa_n}$, and $k > 0$ is some fixed integer that is
prime to $q$ (we may take $k = 1$, for example).  Letting $w \eqdf
\wt(x)$ and $b \eqdf b_n/a_n$ and $c \eqdf c_n/a_n$, we have
\[ U\ket{\psi_{x,L}} = \sum_{j=0}^{q-1} c_j e^{-i\pi
k[(w+j)^2+b(w+j)+c]/q} \ket{(x1^j0^{q-1-j})_L}. \]
We decode this state using $\adj{E}$ to obtain the state
$\ket{x}\ket{\Psi_w}$, where
\[ \ket{\Psi_w} \eqdf \sum_{j=0}^{q-1} c_j e^{-i\pi
k[(w+j)^2+b(w+j)+c]/q} \ket{1^j0^{q-1-j}} \]
is the state of the $q-1$ ancilla qubits.

To see that we have isolated the value $w \bmod q$ in the ancill\ae,
we need only check that, for any integer $v$, \
$\braket{\Psi_v}{\Psi_w} = 0$ if $v \not\equiv w \pmod{q}$.  Note that
all the states of the form $\ket{\Psi_u}$ lie in a $q$-dimensional
subspace $\cH''$ of $\cH$, spanned by $\setof{\ket{1^j0^{q-1-j}} \mid
0\leq j < q}$.  Let $v$ be the Hamming weight of $z$.  Then we have
\begin{eqnarray*}
\braket{\Psi_v}{\Psi_w} & = & \sum_{j=0}^{q-1} |c_j|^2 \exp\left(i\pi
k[(v+j)^2+b(v+j)+c - (w+j)^2-b(w+j)-c]/q\right) \\
& = & q^{-1} \sum_{j=1}^{q-1} \exp\left(i\pi k[v^2 - w^2 + 2j(v-w) +
b(v-w)]/q\right) \\
& \propto & q^{-1} \sum_{j=1}^{q-1} \exp\left(2i\pi jk(v-w)/q\right)
\\
& = & \delta_{(v \bmod q),(w \bmod q)},
\end{eqnarray*}
where $\delta_{x,y}$ is the Kronecker delta.  Thus there is an
orthonormal basis $\setof{\ket{\alpha_j} \mid 0\leq j < q}$ for
$\cH''$ such that, for all integers $w\geq 0$, there are real values
$\theta_w$ such that $\ket{\Psi_w} = e^{i\theta_w}\ket{\alpha_{w \bmod
q}}$.

To finish the simulation, we apply to the ancill\ae\ some (any) operator
$R$ that maps $\ket{\alpha_j}$ to $\ket{1^j0^{q-1-j}}$.  We then use a
CNOT to copy the $j$th ancilla into the target qubit $t_j$.  We then
undo all the previous computations to get rid of any conditional phase
factors.  As was remarked in Section~\ref{sec:heisenberg}, the reverse
computation uses $\adj{U}$, which can be simulated exactly by evolving
via $G_n$ for a positive time only with certain restrictions on the
value of $b$ (the value $c$ is unimportant in that it only results in
an overall phase factor).  If $q \geq 3$, then it is easy to check
that for real $u>0$, \ $\exp(-iuG_n) \propto I$ (restricted
to the space of encoded vectors) if and only if both
$\frac{ua_n(1+b)}{2\pi}$ and $\frac{ua_n(2+b)}{\pi}$ are both
integers.  The latter conditions hold iff $b$ is rational.

Finally, we note that we may be able to get by with less than the full
use of $\adj{E}$ and $E$ on the inside of
$U$ and $\adj{U}$ in the circuit of Figure~\ref{fig:mod-circuit}.  If
the encoded state after applying $U$ is not completely entangled, we
need only decode the ancill\ae\ and those qubits that are entangle
with the ancill\ae, as was done in the circuit of
Figure~\ref{fig:parity-circuit}.

\section{Further Research}

We have assumed throughout that the coupling coefficients $J_{i,j}$ of
(\ref{eqn:heisenberg}) are all equal.  Whether this assumption is
realistic remains to be seen.  It is certainly more likely in the
short run that in feasible laboratory setups, the $J_{i,j}$ will not
be equal, but can still satisfy certain symmetries.  For example, if
$n$ identical spin-$1/2$ particles are arranged in a circular ring, we
would expect the Hamiltonian to be cyclically symmetric, i.e.,
$J_{i,j}$ to depend only on $(i-j) \bmod n$.  For another example, if
the particles are arranged on points in a two- or three-dimensional
regular lattice, we would expect translational symmetry of the
$J_{i,j}$.\footnote{Heisenberg interactions on one-dimensional spin
chains are widely studied.  It is unlikely, however, that these
configurations can be used for parity/fanout, since there are only a
linear number of significant terms in the Hamiltonian, and it can be
shown that a quadratic term (in $j$) in the Hamiltonian is necessary
for our results.}  Computing parity in these more realistic situations
would be very useful and deserving of further investigation.

Heisenberg interactions also figure prominently in recent proposals for
fault-tolerant quantum computation in decoherence-free subspaces (see,
for example, \cite{LCW:DFS,BKLW:DFS} and references cited therein).  The
use of these interactions for this purpose does not appear consistent with
our use here, yet it would be helpful to integrate these two approaches,
perhaps by encoding logical qubits in a DFS.

\section*{Acknowledgments}

We thank Isaac Chuang for first posing the question that eventually
gave rise to the current results: whether spin-exchange interactions
have any use for fast quantum computation.  We also thank Frederic
Green and Steven Homer for interesting and helpful discussions on this
and many other topics.


\end{document}